\documentstyle[epsfig,12pt]{article}
\def\be{\begin{equation}}
\def\ee{\end{equation}}
\def\bea{\begin{eqnarray}}
\def\eea{\end{eqnarray}}

\begin{document}

\title{\Large \bf On the role of the final state interactions in rare B-decays }
\author{\large A.B. Kaidalov, M.I. Vysotsky\bigskip \\
{\it  Institute of Theoretical and Experimental
Physics, Moscow}}

\maketitle


\begin{center}
{\bf Abstract}\\
\medskip
The effects of final state interactions (FSI) in hadronic B-decays
are investigated. The model for FSI, based on Regge
phenomenology of high-energy hadronic interactions is proposed.
It is shown that this model explains the pattern of phases in
matrix elements of $B\to\pi\pi$ and $B\to\rho\rho$ decays. These
phases play an important role for CP-violation  in B-decays.
The most precise determination of the unitarity triangle angle
$\alpha$ from $B_d\to \rho\pi$ decays is
performed. The relation between CP-asymmetries in $B\to K\pi$ decays
is discussed.
It is emphasized that the large distance FSI can explain the structure of
polarizations of the vector mesons in B-decays and other puzzles
like a very large branching ratio of the B-decay to
$\bar\Xi_c\Lambda_c$.
\end{center}

\bigskip


\section{Introduction} \label{s1}
In this paper we give a review of some unusual properties of the matrix
elements in the hadronic B-decays. It is based on papers
\cite{Vys1,Vys2}, where $B\to\pi\pi$, $B\to\rho\rho$ decays were
discussed and it contains some new material on $B\to \rho\pi$, $B\to
K\pi$ decays and polarization of vector mesons in B-decays.
The detailed information on B-decays, obtained in the experiments at
B-factories \cite{hfag}, provides a testing ground for theoretical
models. The investigation of rare B-decays and CP violation in these
decays provides not only the information on CKM matrix, but also on
QCD dynamics both at small and large distances.


One of the most interesting and unsolved problems in
B-decays is the role of FSI. In this paper we shall demonstrate
that FSI play an important role in the hadronic B-decays and enable
to explain some puzzles observed in rare B-decays. In particular
it will be demonstrated that the phases due to strong interactions are
substantial in some hadronic B-decays. These phases are important
for understanding the pattern of CP-violation in rare B-decays.
The model for calculation of FSI will be formulated and compared
to the data on $B\to\pi\pi$ and $B\to\rho\rho$ decays. The model
is based on Regge-picture for high-energy binary amplitudes and
enables to explain a pattern of helicity amplitudes  in
some B-decays to vector mesons. The large distance interactions
provide a simple explanation of the anomalously large branching ratio
of the B-decay to $\bar\Xi_c\Lambda_c$. The CP-violation
asymmetries will be discussed and the most accurate determination of
the unitarity triangle angle $\alpha$ will be presented.

\section{$B\to\pi\pi$/$B\to\rho\rho$ puzzle} \label{s2}

The probabilities of three $B\to\pi\pi$ and three $B\to\rho\rho$
decays are measured now with a good accuracy and  presented in
Table I. There is a large difference between the ratios of the
charged averaged $B_d$ decay probabilities to the charged and
neutral mesons:
\begin{equation}
R_\rho \equiv \frac{\rm Br(B_d \to\rho^+\rho^-)}{\rm Br(B_d
\to\rho^0\rho^0)} \approx 35 \; , \;\; R_\pi \equiv \frac{\rm
Br(B_d \to\pi^+\pi^-)}{\rm Br(B_d \to\pi^0\pi^0)} \approx 4 \;\; .
\label{1}
\end{equation}

It was demonstrated in refs.\cite{Vys1,Vys2} that this difference
is related to the difference of phases due to strong interactions
for matrix elements of $B\to\pi\pi$ and $B\to\rho\rho$-decays. The
matrix elements of these decays can be expressed in terms of
amplitudes with isospin zero and two. To take into account the
differences in CKM phases for
tree and penguin contributions we separate the amplitude with I=0
into the corresponding parts $A_0$ and $P$:
\begin{eqnarray}
M_{\bar B_d \to \pi^+\pi^-} & = & \frac{G_F}{\sqrt 2}|V_{ub}
V_{ud}^*| m_B^2 f_\pi f_+(0) \left\{e^{-i\gamma}\frac{1}{2\sqrt
3}A_2 e^{i\delta_2^\pi} \right. + \nonumber \\ & + & e^{-i\gamma}
\left.\frac{1}{\sqrt 6} A_0 e^{i\delta_0^\pi} +
\left|\frac{V_{td}^* V_{tb}}{V_{ub} V_{ud}^*}\right| e^{i\beta} P
e^{i(\delta_P^\pi + \tilde\delta_0^\pi)}\right\} \;\; , \label{2}
\end{eqnarray}
\begin{eqnarray}
M_{\bar B_d \to \pi^0\pi^0} & = & \frac{G_F}{\sqrt 2}|V_{ub}
V_{ud}^*| m_B^2 f_\pi f_+(0) \left\{e^{-i\gamma}\frac{1}{\sqrt
3}A_2 e^{i\delta_2^\pi} \right. - \nonumber \\ & - & e^{-i\gamma}
\left.\frac{1}{\sqrt 6} A_0 e^{i\delta_0^\pi} -
\left|\frac{V_{td}^* V_{tb}}{V_{ub} V_{ud}^*}\right| e^{i\beta} P
e^{i(\delta_P^\pi  + \tilde\delta_0^\pi)}\right\} \;\; , \label{3}
\end{eqnarray}
\begin{equation}
M_{\bar B_u \to \pi^-\pi^0} = \frac{G_F}{\sqrt 2}|V_{ub} V_{ud}^*|
m_B^2 f_\pi f_+(0) \left\{\frac{\sqrt 3}{2\sqrt 2} e^{-i\gamma}
A_2 e^{i\delta_2^\pi} \right\} \;\; , \label{4}
\end{equation}
where $V_{ik}$ are the elements of CKM matrix, $\gamma$ and
$\beta$ are the unitarity triangle angles and we factor out the
product $m_B^2 f_\pi f_+(0)$ which appears when the decay
amplitudes are calculated in the factorization approximation.

The charge conjugate amplitudes are obtained by the same formulas
with substitution $\beta, \gamma \to -\beta, -\gamma$.

The CP asymmetries are given by \cite{G}: $$ C_{\pi\pi} \equiv
\frac{1-|\lambda_{\pi\pi}|^2}{1+|\lambda_{\pi\pi}|^2} \; , \;\;
S_{\pi\pi} \equiv \frac{2{\rm Im}(\lambda_{\pi\pi})}{1+
|\lambda_{\pi\pi}|^2} \; , \;\; \lambda_{\pi\pi} \equiv
e^{-2i\beta} \frac{M_{\bar B \to \pi\pi}}{M_{B\to\pi\pi}} \;\; ,
$$ where $\pi\pi$ is $\pi^+\pi^-$ or $\pi^0 \pi^0$.

The analogous formulas take place for $\rho\rho$ final states where
the longitudinal polarizations of $\rho$-mesons are dominant.

The values of $P$ can be determined using $d \leftrightarrow s$ interchange symmetry
from decays $B_u\to K^{0*}\rho^+$ and $B_u\to K^0 \pi^+$
\cite{Gronau} and turn out to be rather small compared to tree
contributions. Note, however, that $P$ determines the magnitudes of
the direct CP violation in hadronic decays. 

If we neglect the penguin
contribution, then the difference of phases is expressed in terms
of the branching ratios as follows
\begin{equation}
\cos(\delta_0^\pi -\delta_2^\pi) = \frac{\sqrt 3}{4} \frac{{\rm
B}_{+-} - 2 B_{00} + \frac{2}{3} \frac{\tau_0}{\tau_+}
B_{+0}}{\sqrt{\frac{\tau_0}{\tau_+} B_{+0}}\sqrt{B_{+-} + B_{00}
-\frac{2}{3} \frac{\tau_0}{\tau_+} B_{+0}}} \;\; . \label{8}
\end{equation}
Using the experimental information on the branching ratios of
$B\to\pi\pi$-decays \cite{hfag} we obtain $|\delta_0^\pi -
\delta_2^\pi| = 48^o$.

The penguin contributions to $B_{ik}$ do not interfere with the tree ones
because CKM angle $\alpha = \pi -\beta -\gamma$ is almost equal to
$\pi/2$. Taking into account  P-term we get:
\begin{equation}
|\delta_0^\pi - \delta_2^\pi| = 37^o \pm 10^o \;\; . \label{9}
\end{equation}
This agrees with the result of the analysis in ref.\cite{Cheng}:
\begin{equation}
\delta_0^\pi - \delta_2^\pi = 40^o \pm 7^o \;\; . \label{11}
\end{equation}
Thus the difference of the phases of the matrix elements with I=0 and I=2
is not small in sharp contrast with the factorization approximation
often used for estimates of heavy meson decays.

For  $B\to\rho\rho$-decays we obtain in the analogous way:
\begin{equation}
|\delta_0^\rho - \delta_2^\rho| = {11^o}^{+6^o}_{-11^o} \;\; .
\label{12}
\end{equation}
This phase difference is smaller than for pions and is consistent
with zero.

The fact that the phases due to FSI are in general not small for heavy
quark decays is confirmed by the other D and B-decays. The data on
$D\to\pi^+ \pi^-$, $D \to \pi^0 \pi^0$ and $D^\pm \to \pi^\pm
\pi^0$ branching ratios lead to \cite{Cl}:
\begin{equation}
\delta_2^D - \delta_0^D \equiv \delta_D = \pm(86^o \pm 4^o) \;\; .
\label{30}
\end{equation}
The last example is $B\to D\pi$ decays. $D\pi$ pair produced in
$B$-decays can have $I=1/2$ or $3/2$. From the measurement of the
probabilities of $B^- \to D^0 \pi^-$, $B^0 \to D^- \pi^+$ and $B^0
\to D^0 \pi^0$ decays in paper \cite{Cleo} the FSI phase
difference of these two amplitudes was determined:
\begin{equation}
\delta_{D\pi} = 29^o \pm 4^o \;\; . \label{34}
\end{equation}
   Thus the experimental data indicate that the phases due to FSI
are not small for heavy meson decays.

\section{Calculation of the FSI phases of $B\to\pi\pi$
and $B\to\rho\rho$ decay amplitudes} \label{s3}

 Let us remind that for $K\to\pi\pi$ decays there are no inelastic
 channels, Migdal-Watson (MW) theorem is applicable and strong
 interaction
 phases of $S$-matrix elements of $K\to (2\pi)_I$ decays are equal to
 the phases of the corresponding $\pi\pi \to \pi\pi$ scattering
 amplitudes at $E = m_K$.

 For B-mesons there are many opened inelastic channels and MW
 theorem is not directly applicable. Serious arguments that strong
 phases should disappear in the $M_Q\to\infty$ limit were given by
 J.D. Bjorken \cite{BJ}. He emphasized the fact that the
 characteristic configurations of the  light quarks
produced in the decay  have small size $\sim 1/M_Q$ and FSI interaction cross
 sections should decrease as $ 1/M^2_Q$. Similar arguments were
 applied in the analysis of heavy quark decays in the QCD
 perturbation theory \cite{Noibert}. These arguments can be
 applied to the total hadronic decay rates. For individual decay
 channels (like $B\to\pi\pi$) which are suppressed in the limit
 $M_Q\to\infty$ the situation is more delicate. However, even in
 these situations the arguments of Bjorken that due to large
 formation times the final particles are formed and can interact
 only at large distances from the point of the decay seem
 relevant.

 On the other hand, the formal analysis of different classes of
 Feynman diagrams, including soft rescatterings \cite{Kaid,Don},
 show that the diagrams with pomeron exchange in the
 FSI-amplitudes do not decrease as $M_Q$ increases. The same
 conclusions follow from the applications of generalizations of
 MW-theorem \cite{Wolf,Deandr}.

 In the process of the analysis of FSI in heavy meson decays it is
 important to understand the structure of the intermediate
 multiparticle states. It was shown in ref.\cite{Vys2} that the
 bulk of multiparticle states produced in heavy meson decays has
 a small probability to transform into two-meson final state and
 only quasi two-particle intermediate states $XY$ with the masses
 $M^2_{X(Y)}\leq M_B \Lambda_{QCD}\ll M^2_B$ can be effectively
 transformed into the final two-meson state. In
 refs.\cite{Vys1,Vys2}
 in calculation of FSI effects for $B\to\pi\pi$ and $B\to\rho\rho$
 decays
  only two particle intermediate states with positive
 $G$-parity to which $B$-mesons have relatively large decay
probabilities
 were considered. Alongside with $\pi\pi$ and $\rho\rho$ there is 
only one such
state: $\pi a_1$.

We shall use Feynman diagrams approach to calculate FSI phases
from the diagram with the low mass intermediate states $X$ and
$Y$. Integrating over loop momenta $d^4 k$ one can transform the
integral over $k_0$ and $k_z$ into the integral over the invariant
masses of clusters of intermediate particles $X$ and $Y$:
\begin{equation}
\int dk_0dk_z=\frac{1}{2M_B^2}\int ds_X ds_Y \;\; ,\label{260}
\end{equation}
and deform integration contours in such a way that only the low mass
intermediate states contributions are taken into account while the
contribution of heavy states being small is neglected. In this way
we get:
\begin{equation}
M_{\pi\pi}^I = M_{XY}^{(0)I} (\delta_{\pi X}\delta_{\pi Y} + i
T_{XY \to\pi\pi}^{J=0}) \;\; , \label{26}
\end{equation}
where $M_{XY}^{(0)I}$ are the decay matrix elements without FSI
interactions and $T_{XY\to\pi\pi}^{J=0}$ is the $J=0$ partial wave
amplitude of the process $XY\to\pi\pi \;\; (T^J = (S^J -1)/(2i))$
which originates from the integral over $d^2k_{\perp}$.

For real $T$  Eq.(12) coincides with the application of the
unitarity condition for the calculation of the imaginary part of
$M$ while for the imaginary $T$ the corrections to the real part
of $M$ are generated.

This approach is analogous to the FSI calculations performed in
paper \cite{111}. In \cite{111} $2 \rightarrow 2$ scattering
amplitudes were considered to be due to elementary particle
exchanges in the $t$-channel. For vector particle exchanges
$s$-channel partial wave amplitudes behave as $s^{J-1} \sim s^0$
and thus do not decrease with energy (decaying meson mass).
However it is well known that  the correct behavior is given by
Regge theory: $s^{\alpha_i(0)-1}$. For $\rho$-exchange
$\alpha_{\rho}(0) \approx 1/2$ and the amplitude decrease with
energy as $1/\sqrt s $. This effect is very spectacular for $B\to
DD\to\pi\pi$ chain with $D^*(D^*_2)$ exchange in $t$-channel:
$\alpha_{D^*}(0) \approx -1$ and reggeized $D^*$ meson exchange is
damped as $s^{-2} \approx 10^{-3}$ in comparison with the elementary
$D^*$ exchange (see for example \cite{Deandr}). For
$\pi$-exchange, which gives a dominant contribution to $\rho\rho
\to \pi\pi$ transition (see below), in the small $t$ region the
pion is close to mass shell and its reggeization is not important.

Note that the pomeron contribution does not decrease for
$M_Q\to\infty$, however it does not contribute to the difference
of phases $\delta_0^\pi - \delta_2^\pi$ which we are interested
in. So this phase difference is determined by the secondary
exchanges ($\rho,\pi$) and it decreases at least as $1/M_Q$ for
large $M_Q$ in accordance with Bjorken arguments. For phases
$\delta_0^\pi$ and $ \delta_2^\pi$ separately the pomeron
contribution does not cancel in general. If Bjorken arguments are
valid for these quantities it can happen only under exact
cancellation of different diffractively produced intermediate
states and it does not happen in the model of
refs.\cite{Vys1,Vys2}.

Let us calculate the imaginary parts of $B\to\pi\pi$ decay
amplitudes which originate from $B\to\rho\rho\to\pi\pi$ chain :
\begin{equation}
{\rm Im} M(B\to\pi\pi) = \int\frac{d\cos\theta}{32\pi}
M(\rho\rho\to\pi\pi) M^* (B\to\rho\rho) \;\; . \label{13}
\end{equation}
 In the amplitude $\rho\rho\to\pi\pi$ of $\rho\rho$ intermediate
state the exchange by pion trajectory in the t-channel
dominates. It was already  stressed that $\rho$-mesons produced in
$B$-decays are almost entirely longitudinally polarized. That is
why it is necessary to take into account only longitudinal
polarizations for the intermediate $\rho$-mesons. The amplitude of
$\rho^+\rho^0\to\pi^0\pi^+$ transition is determined by the well
known constant $g_{\rho\to\pi\pi}$. This contribution is the
dominant one for $B\to\pi\pi$ decays due to a large probability of
$B\to\rho\rho$-transition. Let us note that in the limit $M_B
\to\infty$ the ratio $Br(B_d\to\rho\rho)/Br(B_d\to\pi\pi)$ grows
as $M^2_B$, that is why  FSI phase $\delta_2^\pi(\rho\rho)$  (and
$\delta_0^\pi(\rho\rho)$) diminishes only as $1/M_B$. On the contrary
$\pi\pi$ intermediate state plays a minor role in
$B\to\rho\rho$-decays.

In description of $\pi\pi$ elastic scattering amplitudes in
Eq.(12) the contributions of $P,f$ and $\rho$ Regge-poles were taken
from ref.\cite{Bor}. Finally $\pi a_1$ intermediate state should
be taken into account. The large branching ratio of $B_d\to\pi^{\pm}
a_1^{\mp}$-decay ( ${\rm Br}(B_d\to\pi^{\pm} a_1^{\mp})=(40 \pm
4)*10^{-6}$) is partially compensated by the small $\rho\pi a_1$
coupling constant (it is $1/3$ of $\rho\pi \pi$ one). As a result
the contribution of  $\pi a_1$ intermediate state (which
transforms into $\pi\pi$ by $\rho$-trajectory exchange in the
$t$-channel) to FSI phases equals approximately that part of
$\pi\pi$ intermediate state contribution which is due to
$\rho$-trajectory exchange. Assuming that the sign of the $\pi
a_1$ intermediate state contribution to phases is the same as
that of the elastic channel and taking into account that the loop
corrections to  $B\to\pi\pi$ decay amplitudes lead to the
diminishing of the (real) tree amplitudes by $\approx 30\%$ we
obtain:
\begin{equation}
\delta_0^\pi = 30^o \; , \;\;\delta_2^\pi = -10^o\;\;,\;\;
\delta_0^\pi - \delta_2^\pi = 40^o \;\; . \label{263}
\end{equation}
The accuracy of this prediction is about $15^o$.

 For $\rho\rho$ final state the analogous difference is about three
times smaller,
$\delta_0^\rho - \delta_2^\rho \approx 15^o $. Thus the proposed
model for FSI enables us to explain the $B\to\pi\pi$/$B\to\rho\rho$
puzzle.

\section{Direct CPV in $B\to\pi\pi$-decays and phases of the penguin contribution}\label{s4}

It follows from Eq. (2)  that the direct CP asymmetry
 in $B_d(\bar B_d) \to \pi^+\pi^-$ decay has the following
expression in terms of quantities $A_0,A_2,P$ and phases:
\begin{eqnarray}
C_{+-} & = & -\frac{\tilde P}{\sqrt 3} \sin\alpha [\sqrt 2 A_0
\sin(\delta_0 - \tilde\delta_0 - \delta_P) + A_2 \sin(\delta_2 -
\tilde\delta_0 -\delta_P)]/ \nonumber \\ & / &[\frac{A_0^2}{6} +
\frac{A_2^2}{12} + \frac{A_0 A_2}{3\sqrt 2} \cos(\delta_0 -
\delta_2) - \sqrt{\frac{2}{3}} A_0 \tilde P \cos\alpha
\cos(\delta_0 - \tilde\delta_0 - \delta_P) - \nonumber \\ & - &
\frac{A_2 \tilde P}{\sqrt 3} \cos\alpha \cos(\delta_2 -
\tilde\delta_0 - \delta_P) + \tilde P^2] \;\; , \label{264}
\end{eqnarray}

where
\begin{equation}
\tilde P \equiv \left|\frac{V_{td}^* V_{tb}}{V_{ub}
V_{ud}^*}\right| P \;\; . \label{265}
\end{equation}

Thus the direct CP-violation parameter 
is proportional to the modulus of the penguin amplitude and is
sensitive to the difference of the strong phases of $A_0, A_2$ and
penguin amplitudes. So far we have discussed the phases of the
amplitudes $A_0, A_2$. The penguin diagram contains a c-quark loop
and has a nonzero phase even in the QCD perturbation theory. It
was estimated in ref.\cite{Vys1} and is about $10^o$. Note that in
PQCD it has a positive sign.

Let us estimate the phase of the penguin amplitude
$\delta^{\pi}_P$ considering the charmed mesons intermediate states:
$B\to \bar D D, \bar{D^*} D, \bar D D^*, \bar{D^*} D^* \to
\pi\pi$. In Regge model all these amplitudes are described at high
energies by the exchanges of $D^*(D^*_2)$-trajectories. An intercept
of these exchange-degenerate trajectories can be obtained using
the method of \cite{k} or from the masses of $D^*(2007) \; $--$\; 1^-$
and $D^*_2(2460) \;$--$\; 2^+$ resonances, assuming linearity of
these Regge-trajectories. Both methods give $\alpha_{D^*}(0) =
-0.8 \div -1$ and the slope $ \alpha_{D^*}' \approx 0.5 GeV^{-2}$.

The amplitude of $D^+D^-\to\pi^+\pi^-$ reaction in the Regge model
proposed in paper \cite{b} can be written in the following form:
\begin{equation}
T_{D\bar D \to\pi\pi}(s,t)=-\frac{g^2_0}{2}e^{-i\pi\alpha(t)}
\Gamma(1-\alpha_{D^*}(t))(s/s_{cd})^{\alpha_{D^*}(t)} \;\; ,
\label{1259}
\end{equation}
where $\Gamma (x)$ is the gamma function.

The $t$-dependence of Regge-residues is chosen in accordance with the
dual models and is tested for light (u,d,s) quarks. According to
\cite{b} $s_{cd} \approx 2.2~ GeV^2$.

Note that the sign of the amplitude is fixed by the unitarity in
the $t$-channel (close to the $D^*$-resonance). The constant
$g^2_0$ is determined by the width of the $D^* \to D \pi$ decay:
$g^2_0/(16\pi)=6.6$. Using eq.(9) and the branching ratio $Br
(B\to D \bar D)\approx 2\cdot 10^{-4}$ we obtain the imaginary
part of $P$ and comparing it with the contribution of $P$ in $B\to
\pi^+\pi^-$ decay probability we get $\delta^{\pi}_P \approx
-3.5^o$. The sign of $\delta_P$ is negative - opposite to the
positive sign which was obtained in perturbation theory. Since
$D\bar D$-decay channel constitutes only $\approx 10\%$ of all
two-body charm-anticharm decays of $B_d$-meson, taking these
channels into account we easily get
\begin{equation}
\delta_P \sim -10^o \;\; , \label{1260}
\end{equation}
which may be very important for the interpretation of the
experimental data on direct CP asymmetry.

It was shown in ref.\cite{Vys2} that assuming that the phases satisfy
the conditions: $\delta_0 - \delta_2=37^o,  \delta_2\leq 0$ and
$\delta_P> 0$, it is possible to obtain the following inequality
\begin{equation}
C_{+-} > -0.18 \; .\label{270}
\end{equation}

It is worthwhile to compare the obtained numbers with the value
of $C_{+-}$ which follows from the asymmetry $A_{CP}(K^+ \pi^-)$
if $d \leftrightarrow s$ symmetry is supposed \cite{13a}:

\begin{eqnarray}
C_{+-} & = & \left(\frac{f_{\pi}}{f_K}\right)^2 A_{CP}(K^+ \pi^-)
\frac{\Gamma(B\to K^+ \pi^-)}{\Gamma(B\to \pi^+ \pi^-)} \frac{
\sin(\beta +
\gamma)}{\sin(\gamma)}\left|\frac{V_{td}}{V_{ts}\lambda}
\right| = \nonumber \\
& = & 1.2^{(-2)}(-0.093 \pm 0.015)\frac{19.8}{5.2}
\frac{\sin82^o}{\sin60^o} 0.87=-0.24 \pm 0.04 \;\; .\label{271}
\end{eqnarray}

Experimental results obtained by Belle \cite{14} and BABAR
\cite{15} are contradictory:
\begin{equation}
C_{+-}^{Belle}=-0.55(0.09) \;\; , C_{+-}^{BABAR}=-0.21(0.09),
\label{272}
\end{equation}
with Belle number being far below (\ref{270}). For a non-perturbative
phase of the penguin contribution (\ref{1260}) the value of the
theoretical prediction for $C_{+-}$ can be made substantially
smaller and closer to the Belle result.

For direct CP asymmetry in $B_d(\bar B_d)\to \pi^0\pi^0$ decay
from (\ref{3}) we readily obtain:

\begin{eqnarray}
C_{00} & = & -\sqrt{\frac{2}{3}} \tilde P \sin\alpha [A_0
\sin(\delta_0 - \tilde\delta_0 -\delta_P) - \sqrt 2 A_2
\sin(\delta_2 - \tilde\delta_0 -\delta_P)] / \nonumber \\
& / & [\frac{A_0^2}{6} + \frac{A_2^2}{3} - \frac{\sqrt 2}{3} A_0
A_2 \cos(\delta_0 - \delta_2) - \sqrt{\frac{2}{3}} A_0 \tilde P
\cos\alpha \cos(\delta_0 - \tilde\delta_0 - \delta_P) + \nonumber
\\ & + &
\frac{2}{\sqrt 3} A_2 \tilde P \cos\alpha \cos(\delta_2 -
\tilde\delta_0 -\delta_P) + \tilde P^2] \;\; , \label{273}
\end{eqnarray}

\begin{equation}
C_{00} \approx  -1.06[0.8\sin(\delta_0 - \tilde\delta_0 -
\delta_P)- 1.4\sin(\delta_2 - \tilde\delta_0 - \delta_P)] \approx
-0.6 \;\; . \label{274}
\end{equation}

This unusually large direct CPV (measured by $|C_{00}|$) is
intriguing task for future measurements since the present
experimental error is too big:
\begin{equation}
C_{00}^{exper}=-0.48(0.32) \;\; .     \label{275}
\end{equation}

Another  CPV asymmetry measured in $B_d(\bar B_d)\to \pi\pi$
decays $S_{+-}$ is sensitive to the unitarity triangle angle
$\alpha$. Let us first neglect the penguin contribution. Then
from the experimental value $S_{+-}^{exper}= -0.62 \pm 0.09$
\cite{14,15} we get:

\begin{equation}
\sin 2\alpha^{\rm T} = S_{+-} \;\; , \label{276}
\end{equation}

\begin{equation}
\alpha^{\rm T}=109^o \pm 3^o\;\; . \label{277}
\end{equation}

The penguin shifts the value of $\alpha$. The accurate formula looks
like:
\begin{eqnarray}
S_{+-} & = &[\sin 2\alpha(\frac{A^2_0}{6} + \frac{A^2_2}{12} +
\frac{A_0 A_2}{3\sqrt 2}\cos(\delta_0 - \delta_2)) - \nonumber \\
& - &\frac{A_2 \tilde P}{\sqrt 3} \sin\alpha \cos(\delta_2 -
\tilde\delta_0 - \delta_P) - \sqrt{\frac{2}{3}} A_0 \tilde P
\sin\alpha \cos(\delta_0 - \tilde\delta_0 - \delta_P)] / \nonumber
\\ & / & [\frac{A_0^2}{6} +
\frac{A_2^2}{12} + \frac{A_0 A_2}{3\sqrt 2} \cos(\delta_0 -
\delta_2) - \sqrt{\frac{2}{3}} A_0 \tilde P \cos\alpha
\cos(\delta_0 - \tilde\delta_0 - \delta_P) - \nonumber \\ & - &
\frac{A_2 \tilde P}{\sqrt 3} \cos\alpha \cos(\delta_2 -
\tilde\delta_0 - \delta_P) + \tilde P^2] \;\; . \label{278}
\end{eqnarray}

The numerical values of $\alpha$ from different B-decays will be given
in the next Section.

\section{Analysis of $B_d(\bar B_d) \to \rho^\pm \pi^\mp$ decays}

The time dependence of these decay
 probabilities are given by
the following formula \cite{G}:

\begin{eqnarray}
&&\frac{dN(B_d (\bar B_d)\to \rho^\pm \pi^\mp)}{d \Delta t} =
\left(1\pm A_{\rm CP}^{\rho\pi}\right) e^{-t/\tau} \times \nonumber \\
&\times& \left[1-q(C_{\rho\pi} \pm \Delta C_{\rho\pi}) \cos(\Delta
mt) + q(S_{\rho\pi}\pm \Delta S_{\rho\pi}) \sin (\Delta mt)
\right] \;\; ,  \label{1}
\end{eqnarray}
where $q=-1$ corresponds to the decay of a particle which was
$B_d$ at $t=0$, while $q=1$ corresponds to the decay of a particle
which was $\bar B_d$ at $t=0$. According to \cite{G}:

\begin{equation}
A_{\rm CP}^{\rho\pi} = \frac{|A^{+-}|^2 - |\bar A^{-+}|^2 + |\bar
A^{+-}|^2 - |A^{-+}|^2}{|A^{+-}|^2 + |\bar A^{-+}|^2 + |\bar
A^{+-}|^2 + |A^{-+}|^2} \;\; , \label{2}
\end{equation}

where $A^{\pm\mp}$ are the amplitudes of $B_d \to\rho^\pm \pi^\mp$
decays, while $\bar A^{\pm\mp}$ are the amplitudes of $\bar B_d
\to \rho^\pm \pi^\mp$ decays. Introducing the ratios of the decay
amplitudes:

\begin{equation}
\lambda^{\pm\mp} = \frac{q}{p} \frac{\bar A^{\pm\mp}}{A^{\pm\mp}}
\;\; , \label{3}
\end{equation}

where $q/p = e^{-2i\beta}$ comes from $B_d - \bar B_d$ mixing and
$\beta$ is the angle of the unitarity triangle, we obtain the
expressions for the remaining parameters entering Eq. (\ref{1}):

\begin{equation}
C_{\rho\pi} \pm \Delta C_{\rho\pi} =
\frac{1-|\lambda^{\pm\mp}|^2}{1+|\lambda^{\pm\mp}|^2} \; , \;\;
S_{\rho\pi} \pm \Delta S_{\rho\pi} = \frac{2 Im
\lambda^{\pm\mp}}{1+ |\lambda^{\pm\mp}|^2} \;\; , \label{4}
\end{equation}
where $C_{\rho\pi}$ and $S_{\rho\pi}$ (as well as $A_{\rm
CP}^{\rho\pi}$) are CP-odd observables, while $\Delta C_{\rho\pi}$
and $\Delta S_{\rho\pi}$ are CP-even. The experimental data for the
observables entering Eq. (\ref{1}) accompanied by the averaged
branching fraction are presented in Table 2 \cite{hfag}.

The decay amplitudes $\bar A^{\pm\mp}$ are described by the tree and
penguin Feynman diagrams shown in Fig.1. The analogous diagrams
describe amplitudes $A^{\pm\mp}$. The corresponding formulas for the
amplitudes look like:

\begin{eqnarray}
\bar A^{-+}  =  A_1 e^{-i\gamma} + P_1 e^{i(\beta + \delta_1)}
\;\; , \nonumber \\
A^{-+} =  A_2 e^{i\gamma} + P_2 e^{-i(\beta - \delta_2)}
\;\; , \nonumber \\
\bar A^{+-}  =  A_2 e^{-i\gamma} + P_2 e^{i(\beta + \delta_2)}
\;\; ,  \nonumber\\
A^{+-}  =  A_1 e^{i\gamma} + P_1 e^{-i(\beta - \delta_1)}
\;\; , \nonumber \\
A_1/A_2 \equiv a_1/a_2 e^{i\tilde\delta} \;\; , \nonumber \\
P_1/P_2  \equiv  p_1/p_2 e^{i\tilde\delta} \;\; ,  \label{5}
\end{eqnarray}
where $\gamma$ and $\beta$ are the angles of the unitarity triangle, while
$\delta_1$ and $\delta_2$ are the difference of FSI strong phases
between penguin and tree amplitudes (for penguin amplitudes we use
the so-called $t$-convention, subtracting charm quark contribution
to penguin amplitudes).

All in all we have seven parameters in Eq.(\ref{5}) specific for
$\rho\pi$ final states ($a_1, a_2, p_1, p_2, \delta_1, \delta_2$
and $\tilde\delta$) plus UT angle $\alpha = \pi -\beta -\gamma$,
while the number of the experimental observables in Table 1 is six. To
go further we should involve additional theoretical information in
order to reduce the number of parameters. If we find the values of
$p_1$ and $p_2$ even with considerable uncertainties it will be
very helpful for determination of UT angle $\alpha$, since penguin
amplitudes shift $\alpha$ by small amount proportional to
$p_i/a_i$, and even large uncertainty in this shift leads to few
degrees (theoretical) uncertainty in $\alpha$ (see below).

The most straightforward way is to calculate the matrix elements of
the corresponding weak interactions Lagrangian with the help of
factorization, as it was done in \cite{51}. However it was shown
above that there are substantial deviations from factorization in
$B\to\pi\pi$ decays. In particular from the experimental data on
direct CP-asymmetry in $B_d(\bar B_d) \to \pi^+ \pi^-$ decays we
know that the factorization strongly underestimates the contribution of a
penguin diagram to the decay amplitude \cite{Vys1,Vys2}. Another
approach is to extract the penguin amplitudes from the branching
ratios of the $B^- \to \bar K^{0*} \pi^+$ and $B^- \to \bar K^0
\rho^+$ decays in which the penguin dominates with the help of
$s\leftrightarrow d$ quark interchange symmetry, analogously to
what was done for penguins in $B\to\pi\pi$ \cite{5} and
$B\to\rho\rho$ \cite{6} decays.

Feynman diagrams responsible for these decays are shown in Fig. 2.
Comparing Fig. 2 with Fig. 1 (b) we readily get the following
relations:

\begin{eqnarray}
Br(\bar B_d \to \pi^+ \rho^-)_{P_1} & = &
\frac{\tau_{B_d}}{\tau_{B_u}} Br(B^- \to \bar K^{0*}\pi^-)
\left|\frac{V_{td}}{V_{ts}}\right|^2
= \\
& = & \frac{1}{1.071}(10.7 \pm 0.8) \cdot 10^{-6} \cdot (0.20)^2 =
0.40(4) \cdot 10^{-6} \;\; , \nonumber \label{6}
\end{eqnarray}

\begin{eqnarray}
Br(\bar B_d \to \rho^+\pi^-)_{P_2} & = &
\frac{\tau_{B_d}}{\tau_{B_u}} Br(B^- \to \bar K^0\rho^-)
\left|\frac{V_{td}}{V_{ts}}\right|^2
= \\
& = & \frac{1}{1.071}(8.0 \pm 1.5) \cdot 10^{-6} \cdot (0.20)^2 =
0.30(6) \cdot 10^{-6} \;\; , \nonumber \label{7}
\end{eqnarray}
from which the values of $p_1$ and $p_2$ follow:

\begin{equation}
p_1^2 = 0.40(4) \cdot 10^{-6} \; , \;\; p_2^2 = 0.30(6) \cdot
10^{-6} \;\; , \label{8}
\end{equation}
where here and below we neglect the common factor $16\pi m_B
\Gamma_{B_d}$, to which squares of amplitudes are proportional.
The remaining $8-2=6$ parameters entering Eq.(\ref{5}) we will determine
from  six experimental numbers presented in Table 2.

From Eq.(\ref{5}) we get the following relation for the averaged
branching ratio of $B_d(\bar B_d)$ decays to $\rho^\pm \pi^\mp$:

\begin{equation}
\frac{a_1^2 + a_2^2}{2} + \frac{p_1^2 + p_2^2}{2} = 23.1(2.7)
\cdot 10^{-6} \;\; , \label{9}
\end{equation}
where the penguin-tree interference terms are omitted (being
proportional to $\cos(\pi - \beta - \gamma) = \cos\alpha$ they are
very small since UT is almost rectangular, $\alpha \approx
\pi/2$).

To determine the values of $a_i$ the equation for $\Delta C_{\rho\pi}$
is helpful:

\begin{equation}
\Delta C_{\rho\pi} = \frac{a_1^2 - a_2^2}{a_1^2 + a_2^2} +
O\left(\frac{p_i^2}{a_i^2}\right) \;\; , \label{10}
\end{equation}
and from (\ref{8}) - (\ref{10}) and the experimental value for $\Delta
C_{\rho\pi}$ from  Table 1 we get:

\begin{equation}
a_1^2 = 31(3) \cdot 10^{-6} \; , \;\; a_2^2 = 14(3) \cdot 10^{-6}
\;\; . \label{11}
\end{equation}

Now from the equations for $C_{\rho\pi}$ and $A_{\rm CP}^{\rho\pi}$
using the experimental data from Table 1 we are able to extract FSI
phases $\delta_1$ and $\delta_2$:

\begin{eqnarray}
C_{\rho\pi} & = & \frac{2p_1 a_1 \sin\delta_1 + 2p_2 a_2
\sin\delta_2}{a_1^2 + a_2^2} + \frac{a_1^2 - a_2^2}{(a_1^2 +
a_2^2)^2} [2p_2 a_2 \sin\delta_2 - 2p_1 a_1 \sin\delta_1] \;\; ,
\nonumber\\
A_{\rm CP}^{\rho\pi} & = & \frac{2 p_1 a_1 \sin\delta_1 - 2 p_2
a_2 \sin\delta_2}{a_1^2 + a_2^2} \;\; ,  \label{12}
\end{eqnarray}

\begin{equation}
\sin\delta_1 = -0.55(30) \; , \;\; \sin\delta_2 = 0.51(40) \;\; ,
\label{13}
\end{equation}
and we see that large experimental errors of $C_{\rho\pi}$ and
$A_{\rm CP}^{\rho\pi}$ do not allow the accurate determination of the
values of FSI phases.

From the equations for $S$ and $\Delta S$ we will determine the values
of $\alpha$ and $\tilde\delta$:

\begin{equation}
S_{\rho\pi} + \Delta S_{\rho\pi}  = \label{14}
\end{equation}
$$ =  2\frac{a_1 a_2
\sin(2\alpha - \tilde\delta)-p_1 a_2 \cos(\delta_1 - \tilde\delta)
- p_2 a_1 \cos(\delta_2 - \tilde\delta) + 2p_1 a_2
\sin\delta_1\sin\tilde\delta}{a_1^2 + a_2^2 + 2p_1 a_1
\sin\delta_1 - 2p_2 a_2 \sin\delta_2} \;\; , $$

\begin{equation}
S_{\rho\pi} - \Delta S_{\rho\pi} = \label{15}
\end{equation}
$$ =  2\frac{a_1 a_2
\sin(2\alpha + \tilde\delta)-p_2 a_1 \cos(\delta_2 + \tilde\delta)
- p_1 a_2 \cos(\delta_1 + \tilde\delta) - 2p_2 a_1
\sin\delta_2\sin\tilde\delta}{a_1^2 + a_2^2 + 2p_2 a_2
\sin\delta_2 - 2p_1 a_1 \sin\delta_1} \;\; , $$
where in (small)
terms proportional to $p_i$ we have substituted $\alpha = \pi/2$.

Substituting the numerical values for the parameters in the denominators we
get:
\begin{equation}
[(6 \pm 4)S_{\rho\pi} - (45 \pm 4)\Delta S_{\rho\pi}] 10^{-6} = 2
a_1 a_2 \sin\tilde\delta \cos 2\alpha \;\; , \label{16}
\end{equation}

\begin{eqnarray}
[(45 \pm 4) S_{\rho\pi} - (6 \pm 4)\Delta S_{\rho\pi}] 10^{-6} =
2a_1 a_2 \sin 2\alpha
\cos\tilde\delta - \nonumber \\
- 2p_2 a_1 \cos(\tilde\delta - \delta_2) - 2p_1 a_2
\cos(\tilde\delta + \delta_1) \;\; . \label{17}
\end{eqnarray}

From the first equation we see that $\tilde\delta$ equals zero or
$\pi$ with $\pm 5^0$ accuracy. For UT angle $\alpha$ from the
second equation neglecting the penguin contributions we obtain:

\begin{equation}
\alpha_{\rho\pi}^T = 90^o \pm 3^o  ({\rm exp}) \;\; , \label{18}
\end{equation}

while taking penguins into account we get:

\begin{equation}
\alpha_{\rho\pi} = 84^o \pm 3^o ({\rm exp}) \;\; , \label{19}
\end{equation}
where $\delta_1 \approx -30^o$ and $\delta_2 \approx 30^o$ were
used.

Thus penguins shift $\alpha$ by 6$^o$ and even assuming 50\%
accuracy of $d\leftrightarrow s$ symmetry which was used to
determine the numerical values of $p_i$ allows us to determine
$\alpha_{\rho\pi}$ with theoretical accuracy which equals the
experimental one, originating from that in $S_{\rho\pi}$ and
pointed out in (\ref{19}):

\begin{equation}
\alpha_{\rho\pi} = 84^o \pm 3^o ({\rm exp}) \pm 3^o({\rm theor})
\;\; . \label{191}
\end{equation}

The consideration of $B_d(\bar B_d)\to\pi\pi$ decays (see Eqs.(25) -
(27)) leads to the following result:
\begin{equation}
\alpha_{\pi\pi} = 88^o \pm 4^o ({\rm exp}) \pm 10^o ({\rm theor})
\;\; , \label{20}
\end{equation}
where a relatively  large theoretical error is due to big (20$^o$)
shift of the tree level value of $\alpha_{\pi\pi}$ by poorly known
penguins and this time (unlike in \cite{Vys2}) we suppose 50\%
theoretical uncertainty in the value of penguin amplitude.

In the case of $B_d (\bar B_d) \to\rho^+ \rho^-$ decays penguin
shifts the value of $\alpha$ by the same amount as is considered
in this paper for $B_d(\bar B_d) \to \rho^\pm \pi^\mp$ decays, so
the theoretical uncertainty is the same:

\begin{equation}
\alpha_{\rho\rho} = 87^o \pm 5^o ({\rm exp}) \pm 3^o ({\rm theor})
\;\; , \label{21}
\end{equation}
while larger experimental uncertainty is due to that in
$S_{\rho\rho}$,

\begin{equation}
S_{\rho\rho} = -0.06 \pm 0.18 \;\; , \label{22}
\end{equation}
which is twice as big as in $S_{\rho\pi}$.

It is interesting to compare the numerical values (\ref{19}),
(\ref{20}), (\ref{21}) with the recent results of the fit of
Unitarity Triangle \cite{7,8}:

\begin{equation}
\alpha^{\rm CKM fitter} = 88^o \pm 6^o \;\; ,\label{23}
\end{equation}

\begin{equation}
\alpha^{\rm UT fit} = 91^o \pm 6^o \;\;. \label{24}
\end{equation}

Large New Physics(NP) contribution to $b\to dg$ penguin could help
to avoid large FSI phases since now the enhancement of direct CPV seen
in $A_{\rm CP}^{\rho\pi}$ will originate from closeness of tree
level and penguin amplitudes. Also puzzle of large $Br B_d (\bar
B_d) \to \pi^0 \pi^0$ can be resolved by NP contribution to
$b\to dg$ penguin comparable with SM one recalculated from $B_u
\to K^0\pi^+$ decay. The bound on such contribution comes from the
coincidence within the errors of the values of $\alpha$ extracted from
$B\to\pi\pi$, $\rho\pi$ and $\rho\rho$ decays, where the penguin
contributions are very different.\footnote{The same argument can
be applied against large NP contributions to $b\to sg$ penguin: if
the same NP does not enhance $b\to dg$ penguin the value of
$\alpha$ from $B\to\pi\pi$ data will be closer to
$\alpha_{\pi\pi}^T = 109^o$ and disagree with that from
$\alpha_{\pi\rho}$ and $\alpha_{\rho\rho}$.}

These are strong arguments in favor of the measurements of the
parameters of $B\to\pi\pi$, $\rho\pi$ and $\rho\rho$ decays with
better accuracy, which can be performed at LHCb and Super B
factory. A search of NP manifestation by different values of UT
angle $\alpha$ extracted from $B\to\pi\pi$ and $B\to\pi\rho,
\rho\rho$ decays is analogous to the one suggested in \cite{9} through
the difference of $\alpha$ extracted from the penguin polluted
$B\to\pi\pi$ decay and from UT analysis based on tree dominated
observables $V_{cb}$ and $\gamma$.

At the end of this section let us note that the results (\ref{20}) and
(\ref{21}) were obtained in the analysis based on isotopic invariance
of strong interactions from the violation of which the additional
uncertainty in $\alpha$ could follow \cite{10}. Fortunately since
in the absence of penguin amplitudes the relation $S_{\pi\pi,
\rho\rho} = \sin 2 \alpha^T$ is free from this type of
uncertainty, it is manifested only as several percent correction to 
the shift of $\alpha$
induced by penguin which is negligible even for
$B\to\pi\pi$ decays.

\section{Direct CPV in $B\to \pi K$ decays}

Recently Belle has published new results of the measurement of CP
asymmetries in $B_d(\bar B_d) \to K^+ \pi^- (K^- \pi^+)$ and
$B^+(B^-) \to K^+ \pi^0 (K^- \pi^0)$ decays \cite{121}:

\begin{equation}
A_{CP}(K^+ \pi^-) \equiv \frac{\Gamma (\bar B_d \to K^- \pi^+) -
\Gamma(B_d \to K^+ \pi^-)}{\Gamma (\bar B_d \to K^- \pi^+) +
\Gamma(B_d \to K^+ \pi^-)} = -0.094(18)(8) \;\; , \label{26K}
\end{equation}

\begin{equation}
A_{CP}(K^+ \pi^0) \equiv \frac{\Gamma (B^- \to K^- \pi^0) -
\Gamma(B^+ \to K^+ \pi^0)}{\Gamma (B^- \to K^- \pi^0) + \Gamma(B^+
\to K^+ \pi^0)} = 0.07(3)(1) \;\; . \label{27K}
\end{equation}

In \cite{121} the 4.5 standard deviations difference of these
asymmetries was considered as a paradox in the framework of the
Standard Model (see also  \cite{1221}) which it really were IF one neglected the color
suppressed tree quark amplitude. Taking into account QCD penguin
diagram and tree diagrams one easily gets the following relation
between CP asymmetries \cite{122}:

\begin{equation}
A_{CP}(K^+ \pi^-) = A_{CP}(K^+ \pi^0) + A_{CP}(K^0 \pi^0) \;\; ,
\label{28K}
\end{equation}

where $A_{CP}(K^0 \pi^0)$ is proportional to the color suppressed tree
amplitude $C$. The experimental value of $A_{CP}(K^0 \pi^0)$ has 
large
uncertainty:

\begin{equation}
A_{CP}(K^0 \pi^0) = -0.14 \pm 0.11 \;\; , \label{29K}
\end{equation}

however with the help of $d\leftrightarrow s$ interchange symmetry
it can be related with CP asymmetry $C_{00}$ of $B_d(\bar B_d) \to
\pi^0 \pi^0$ decays:

\begin{equation}
A_{CP}(K^0 \pi^0) = \frac{\Gamma (B_d \to\pi^0 \pi^0) +\Gamma(\bar
B_d \to \pi^0 \pi^0)}{\Gamma (B_d \to K^0 \pi^0) + \Gamma(\bar B_d
\to \bar K^0 \pi^0)} \left|\frac{V_{us} V_{ts}}{V_{td}}\right|
\frac{\sin\gamma}{\sin\alpha} C_{00} \;\; , \label{30K}
\end{equation}

where the opposite signs in the definitions of $A_{CP}$ and $C_{00}$ are
compensated by a negative sign of $V_{ts}$. The experimental uncertainty
of $C_{00}$ is also very large, that is why we use the above
result (Eq.(23)) for numerical
estimate:

\begin{equation}
C_{00} \approx -0.6 \;\; . \label{31K}
\end{equation}

Substituting (\ref{31K}) in (\ref{30K}) and (\ref{26K}),
(\ref{27K}) and (\ref{30K}) in (\ref{28K}) we finally obtain:

\begin{equation}
-0.094\pm 0.02 = (0.07\pm 0.03)+(-0.07\pm 0.02) \;\; , \label{32K}
\end{equation}

resolving in this way the paradox noted in \cite{121} (the remaining
$\approx 2\sigma$ difference can be safely attributed to
statistical fluctuation). Concluding this Section let us remind
that the absence of color suppression of the tree amplitude of
$B_d \to \pi^0 \pi^0$ decay is explained in Sections 2,3 by large
FSI phases difference of tree amplitudes with isospin zero and
two.

\section{Polarizations of vector mesons in $B\to VV$-decays}

In this Section we consider $B_d(\bar B_d)$ decays into the pair of
light ($\rho, K^*, \varphi$) vector mesons. The short distance
contributions to vector meson production in B-decays lead to the
dominance of the longitudinal polarization of the vector mesons. 
This
is a general property valid in the large $M_Q$- limit due to
helicity conservation for vector currents and corrections should
be $\sim M_V^2/M_Q^2$. It is satisfied experimentally in
$B\to\rho^+\rho^-$ decays, where the contribution of longitudinal
polarization of $\rho$ mesons is $f_L=\Gamma_L/\Gamma=0.968 \pm
0.023$.  Let us note that FSI are not important for these decays;
for example there are no large strong interaction phases generated
by rescattering. The manifestation of this statement is the absence of
enhancement of the color suppressed amplitude which describes the decay
into $\rho^0\rho^0$.

On the other hand there are several B-decays to vector mesons,
where the longitudinal polarizations give only about $50\%$ of decay
rates. For example:\\
for~ $ B^+\to K^{*0}\rho^+ ~~ f_L= 0.48 \pm 0.08, ~B_d\to
K^{*0}\rho^0 ~~f_L= 0.57\pm 0.12, ~ B^+\to \phi K^{*+} ~~f_L= 0.50
\pm 0.07,~B_d\to \phi K^{*0}~~ f_L= 0.491\pm 0.032 $
\cite{hfag}.

This is a real puzzle IF only short distance dynamics  for these
decays is invoked. We would like to argue that strong rescattering
related to large distance dynamics may be responsible for the
observed polarizations pattern. First let us note that in  all the
decays, where $f_L\approx 50\%$, the penguin diagrams give dominant
contribution. In this case a large contribution to the matrix
elements of the decays comes from $D\bar D_s (D^*\bar D_s, D\bar
D^*_s,..)$ intermediate states, which have large branching ratios.
In Section 4 we analyzed $D\bar D$ intermediate state contribution
to a strong phase of the penguin amplitude for $B_d(\bar B_d) \to
\pi\pi$ decays. It was argued that the phase of the order of
10$^o$ can be generated by the charmed mesons intermediate
state.\footnote{Let us emphasize that this considerable contribution
to the subdominant penguin amplitude is not important for the
$B_d(\bar B_d)\to\pi\pi$ widths, where the tree diagram dominates.}
FSI is important due to large branching ratio 
${\rm Br}(B_d\to D \bar
D)\approx 2 \cdot 10^{-4}$ in comparison with
the penguin contribution to
decays to two pions \cite{Vys2}: ${\rm Br}(B\to\pi\pi)_P = 0.6
\cdot 10^{-6}$
extracted with a
help of $d\leftrightarrow s$ symmetry.
In the case of $b\to sg$ penguin dominated decays the
intermediate state contains $DD_s$ pair, and we should compare
${\rm Br}(B_d \to DD_s)\approx 10^{-2}$ with ${\rm Br}(B\to
K^{*0}\rho^+) \approx 10^{-5}$. That is why the relative
contribution of $DD_s$ states are $\sim 2$ times larger than for
$B\to\pi\pi$ penguins. The amplitude of the binary reaction $D\bar
D_s\to VV$ at high energies is dominated by the exchange of
$D^*$-regge trajectory and according to general rules for
spin-structure of regge vertices (see for example \cite{KK}) the final
vector mesons are produced at high energies  transversely
polarized. Thus we expect a large fraction of transverse
polarization of vector mesons in these decays. The value of $f_L$ is
sensitive to intercept of $D^*$-trajectory \cite{Ladissa}. If the
penguin contribution in the decays indicated above is dominant in
the SU(3) limit we have:

\begin{equation}
Br(\phi K^{*0})=Br(K^{*0}\rho^+)=Br(\phi
K^{*+})=2Br(K^{*0}\rho^0) \label{31KK}
\end{equation}

and $f_L$ in all these decays should
be the same. These predictions agree with experimental data
\cite{hfag}.

\section{ Puzzle of charm-anticharm baryons production}

Large probability of B-decay to $\Lambda_c\bar\Xi_c$ has been
observed recently: $Br(B^+\to \Lambda_c^+\bar\Xi_c^0\sim 10^{-3})$
\cite{hfag}. It is surprisingly large compared to the branching of
B-decay to $\Lambda_c^+\bar p = (2.19\pm 0.8)10^{-5}$. From PQCD
point of view both processes are described by similar diagrams
with the substitution of $u\bar d$ (for $\bar p$) by $c\bar s$ (for
$\bar\Xi_c$) and phase space arguments even favor $\bar
p$-production.

On the other hand from the soft rescatterings point of view the large
probabilities of $\bar D D_s (\bar D^* D_s, \bar D D^*_s,...)$
intermediate states, considered in the previous section, can play
an important role in $B^+\to \Lambda_c^+\bar\Xi_c^0$-decays. For
$\Lambda_c^+\bar p$ final states the corresponding two-meson
intermediate states have smaller branchings and, what is even more
important, have different kinematics. For $\bar D D_s,...$
intermediate states the momentum of these heavy states is not
large ($p\approx 1.8~ GeV$) in B rest frame and all light quarks
($u,d,\bar d,\bar s$) are slow in this frame. The final
$\Lambda_c^+\bar\Xi_c^0$ are also rather slow in the B-rest frame
and thus all quarks have large projections to the wave functions
of the final baryons. On the contrary for $\pi D,\rho D,..$ 
intermediate states in $\Lambda_c^+\bar p$-decays momenta of $\bar
u,d$- quarks in light mesons are large and the projections to the wave
functions of final baryons have extra smallness. The resulting
suppression can be estimated in regge-model of ref.\cite{b} with the
nucleon trajectory exchange in the t-channel and is $\sim 10^{-2}$
in accordance with experimental observation.

\section{ Conclusions}

FSI play an important role in two-body hadronic decays of heavy
mesons. Theoretical estimates with account of the lowest
intermediate states give a satisfactory agreement with the experiment
  and provide the explanation of several puzzles observed in B decays.

\section{Acknowledgments}

 We are grateful to P.N. Kopnin for checking some of our numerical
 results.
 This work was supported in part by the grants: RFBR 06-02-17012,
RFBR 06-02-72041-MNTI, RFBR 07-02-00021, RFBR 08-02-00677a, RFBR
08-02-00494, NSh-4568.2008.2, NSh-4961.2008.2
 and by Russian Agency of Atomic Energy.  \\

\newpage
\begin{center}

{\bf Table 1}

\bigskip

\begin{tabular}{|c|c|c|c|}
\hline Mode & ${\rm Br}(10^{-6})$ &  Mode & ${\rm Br}(10^{-6})$ \\
\hline $B_d \to\pi^+ \pi^-$ & $5.2 \pm 0.2$ & $B_d \to
\rho^+\rho^-$ & $24.2 \pm 3.2$ \\ $B_d \to\pi^0 \pi^0$ & $1.3 \pm
0.2$ & $B_d \to \rho^0\rho^0$ & $0.68 \pm 0.27$ \\ $B_u \to\pi^+
\pi^0$ & $5.7 \pm 0.4$ & $B_u \to \rho^+\rho^0$ & $18.2 \pm 3.0$
\\ \hline
\end{tabular}

\vspace{0.5cm}

$C$-averaged branching ratios of $B\to\pi\pi$ and $B\to\rho\rho$
decays.

\end{center}

\bigskip
\newpage
\begin{center}

{\bf Table 2}

\bigskip

\begin{tabular}{|c|c|c|c|c|c|} \hline
&&&&& \\ $Br B_d(\bar B_d) \to$ & $A_{\rm CP}^{\rho\pi}$ &
$C_{\rho\pi}$ &
$\Delta C_{\rho\pi}$ & $S_{\rho\pi}$ & $\Delta S_{\rho\pi}$ \\
$\to \rho^\pm \pi^\mp$ & & & & & \\ \hline $(23.1 \pm 2.7)10^{-6}$
& $-0.13$ & $0.01$ & $0.37$ & $0.01$ & $-0.04$ \\
& $\pm 0.04$ & $\pm 0.07$ & $\pm 0.08$ & $\pm 0.09$ & $\pm 0.10$
\\
\hline
\end{tabular}

\end{center}

\vspace{0.5cm}

 The experimental values of observables which describe $B_d(\bar
B_d)\to \rho^\pm \pi^\mp$ decays.

\newpage
\begin{center}

         FIGURES
\end{center}
\bigskip
\begin{figure}[!htb]

\centering

\epsfig{file=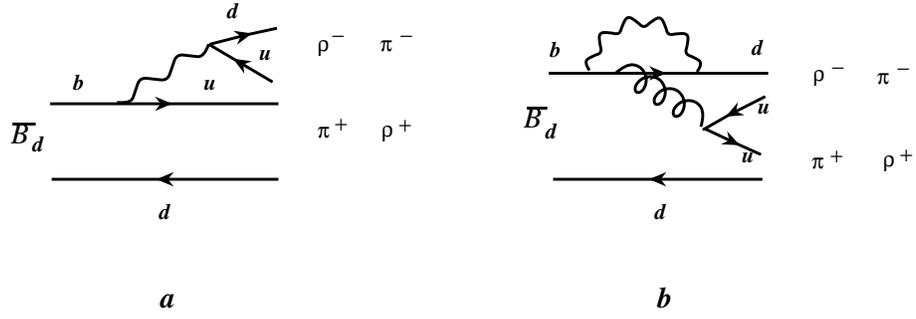,width=12cm}
\bigskip

\caption{ Tree and penguin 
diagrams for B-decays to $\rho\pi$- mesons. $\bar B_d \to
\rho^-\pi^+$ decay is described by the amplitudes $A_1$ and $P_1$,
while  $\bar B_d \to
\rho^+\pi^-$ decay - by the amplitudes $A_2$ and $P_2$.
}
 \label{WW2Fermi}

\end{figure}
\newpage
\begin{figure}[!htb]

\centering

\epsfig{file=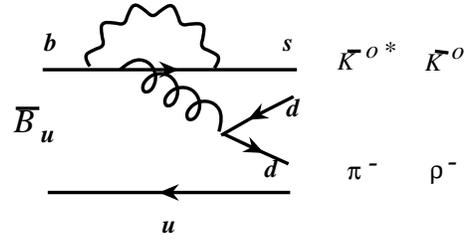,width=6cm}

\caption{  $B^-$ decays in which the penguin diagram dominates. }

\label{WW1Fermi}

\end{figure}


\end{document}